\documentstyle[aps,prb,multicol]{revtex}
\tightenlines
\begin{document}

\include{psfig}

\draft \title{\bf Mixed-state microwave response in superconducting cuprates.}

\author{E. Silva$^{(1)}$\footnote[1]{To whom correspondence should be
addressed.  E-mail: silva@fis.uniroma3.it}, N. Pompeo$^{(2)}$, R.
Marcon$^{(2)}$, R. Fastampa$^{(3)}$, M. Giura$^{(3)}$, S.
Sarti$^{(4)}$, C. Camerlingo $^{(5)}$}

\address{
$^{(1)}$Dipartimento  di Fisica "E. Amaldi" and INFM-Coherentia,
Universit\`a di Roma Tre, Via della Vasca Navale 84, 00146 Roma, Italy\\
$^{(2)}$Dipartimento  di Fisica "E. Amaldi" and INFM,
Universit\`a di Roma Tre, 00146 Roma, Italy\\
$^{(3)}$ Dipartimento di Fisica and INFM-Coherentia, Universit\`a "La Sapienza",
00185 Roma, Italy\\
$^{(4)}$ Dipartimento di Fisica and INFM, Universit\`a "La Sapienza",
00185 Roma, Italy\\
$^{(5)}$ CNR-Istituto di Cibernetica ``E. Caianiello'', 80078 
Pozzuoli, Napoli, Italy}

\date{submitted June 15th, 2004}

\maketitle

\begin{abstract}
We report measurements of the magnetic-field induced microwave complex
resistivity in REBa$_{2}$Cu$_{3}$O$_{7-\delta}$ thin films, with RE =
Y, Sm.  Measurements are performed at 48 GHz by means of a resonant
cavity in the end-wall-replacement configuration.  The magnetic field
dependence is investigated by applying a moderate (0.8 T) magnetic
field along the $c$-axis.  The measured vortex state complex
resistivity in YBa$_{2}$Cu$_{3}$O$_{7-\delta}$ and
SmBa$_{2}$Cu$_{3}$O$_{7-\delta}$ is analyzed within the well-known
models for vortex dynamics.  It is shown that attributing the observed
response to vortex motion alone leads to inconsistencies in the
as-determined vortex parameters (such as the vortex viscosity and the
pinning constant).  By contrast, attributing the entire response to
field-induced pair breaking leads to a nearly quantitative description
of the data.\\


\end{abstract}

\begin{multicols}{2}

\section{Introduction}
\label{intro}
The microwave response is a source of important information in
superconductors.  In high-$T_{c}$ superconductors (HTCS) various
microwave techniques have been used to get information, among the
others, on the symmetry of the order parameter, on the vortex
parameters such as the vortex viscosity and pinning frequency and on the
temperature dependence of the superfluid fraction (via the measurement
of the temperature dependence of the London penetration depth).
In short, microwave measurements are a very powerful tool to study the
superconducting state.\\
While YBa$_{2}$Cu$_{3}$O$_{7-\delta}$ (YBCO) and 
Bi$_{2}$Sr$_{2}$CaCu$_{2}$O$_{8+x}$ (BSCCO) have been the subject of 
intensive experimental investigation at microwave frequencies, other 
HTCS did not receive the same attention.  In particular, a very few 
reports dealt with the vortex-state microwave response in rare-earth 
substituted 123 compounds, mainly on Gd-substituted materials 
\cite{Gd}.  In fact, RE-substituted 123 compounds present interesting 
features in the vortex state: in particular, they often give enhanced 
irreversibility lines and in some cases a slightly higher $T_{c}$.  
Due to their very similarity to YBCO, they are not thought to present 
novel features in the basic mechanisms governing the superconducting 
properties.  Aim of this paper is to present a compared study of the 
microwave response of YBCO and SmBa$_{2}$Cu$_{3}$O$_{7-\delta}$ 
(SmBCO) in the vortex state in moderate fields.  It will be shown that 
some quantitative and qualitative differences exist between the field 
dependence of the complex resistivity in the two materials.  However, 
the analysis of the data shows that those differences cannot be 
related only to changes in the vortex dynamics (e.g., different 
pinning).  Indeed, a close analysis of the data shows that vortex 
motion cannot be the main contribution to the observed resistivity in 
any of the two materials.  Rather, we find that the resistivity of 
both YBCO and SmBCO can be qualitatively described assuming that the 
response is mainly determined by magnetic field induced pair breaking.  
Within this frame, the differences between the two materials can be 
ascribed to different quasiparticle scattering time below $T_{c}$.\\
We recall the main phenomena lying at the base of the microwave
response in the vortex state.  In principle, the response is dictated
by vortex motion, described by the vortex motion complex resistivity
$\tilde\rho_{vm}$, and by charge carriers complex conductivity,
$\tilde\sigma$.  The latter originates from superfluid as well as
normal carriers.\\
A general frame that includes those contributions is the Coffey-Clem 
theory \cite{cc}.  It is noteworthy that the theory, while developed 
for a particular case of vortex motion (vortex in a sinusoidal 
potential submitted to the Lorentz force and to a stochastic force), 
is not constrained to a specific potential shape seen by the vortices.  
Given the vortex motion resistivity 
$\tilde\rho_{vm}$ and the charge carriers complex conductivity 
$\tilde\sigma=\sigma_{1}-\mathrm{i}\sigma_{2}$, the resulting 
expression for the total complex resistivity can be written as:
\begin{equation}
\label{cc}
\tilde\rho=
\frac
{\tilde\rho_{vm}+\frac{\mathrm{i}}{\sigma_{2}}}
{1+\mathrm{i}\frac{\sigma_{1}}{\sigma_{2}}}
\end{equation}
When the field dependence of the pair breaking (affecting
$\tilde\sigma$) is negligible, and for $T$ sufficiently below $T_{c}$
such that $\sigma_{2}\gg\sigma_{1}$, one has for the field induced
change of the resistivity:
\begin{equation}
\label{vm}
\Delta\tilde\rho=\tilde\rho(H,T)-\tilde\rho(0,T)=\tilde\rho_{vm}(H,T)
\end{equation}
since the vortex motion resistivity is zero by definition in zero
field.  By contrast, when the vortex motion resistivity is vanishingly
small, equation \ref{cc} reduces to:
\begin{equation}
\label{2f}
\tilde\rho(H,T)=\frac{1}{\tilde\sigma(H,T)}
\end{equation}
The vortex motion approximation alone has been often used for the 
analysis of the microwave response in the vortex state,\cite{golos} 
sometimes including a field-independent contribution from the 
superfluid.\cite{marcon,revenaz,tsuchiya} In 
this case, from the data one can estimate vortex parameters such as 
the vortex viscosity, the depinning frequency and the pinning constant 
and the activation energy.  By contrast, it has been shown that in 
BSCCO films the imaginary conductivity could be well 
described\cite{mallozzi} by expression \ref{2f}.  In this latter case 
characteristic parameters determining the response are the 
quasiparticle scattering time and the pair-breaking field.

\section{Experimental section}
\label{exp}

\begin{figure}
\centerline{
\psfig{figure=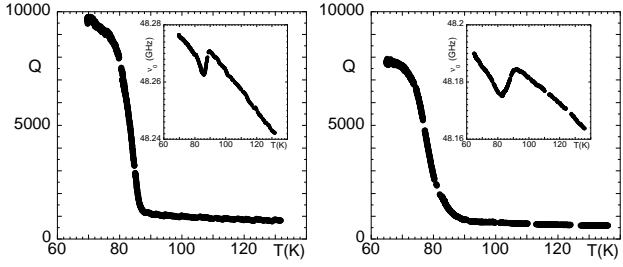,height=3.5cm,clip=,angle=0.}
}
\vspace{1cm}
\caption{Temperature dependence of the coeffcients of the resonant 
cavity incorporating as an end wall the superconducting film. Main panels, 
quality factors; insets, resonant frequency. Left panel: YBCO film; 
right panel, SmBCO film.}
\label{fig1}
\end{figure}

The SmBa$_{2}$Cu$_{3}$O$_{7}$ film was grown on 1 cm $\times$ 1 cm
LaAlO$_{3}$ substrate by planar high oxygen pressure dc sputtering
technique.  The roughness of $\sim$ 20 nm on 1 $\mathrm{\mu}$m
$\times$ 1 $\mathrm{\mu}$m area was measured by AFM. The
YBa$_{2}$Cu$_{3}$O$_{7}$ film was grown by inverted cylindrical
magnetron sputtering on
1 cm $\times$ 1 cm CeO$_{2}$
buffered YSZ
substrates.  In both samples,
out-of-plane
c-axis orientation was assessed by
rocking curves, that showed FWHM $\approx$ 0.2$^{\mathrm{\circ}}$.
Film thickness was $d\sim$200 nm.  $\Phi$-scan showed good
epitaxiality.  YBCO and SmBCO showed $T_{c}\simeq$ 88 K and 87 K,
respectively.  Further details on sample preparation have been
reported elsewhere.\cite{cucolo,camerlingo}\\
The microwave response was measured by the end-wall cavity technique 
at 48.2 GHz in zero dc magnetic field, and with an applied field 
(along the $c$-axis) $\mu_{0}H\leq$0.8 T. The cylindrical cavity
resonated in the TE$_{011}$ mode.\cite{silvaMST}
In this configuration the microwave field probed a circular area, 
corresponding to the diameter of the cavity,
$\sim$ 8 mm, centered on the center of the film.  We note
that in this configuration the edges of the film are not exposed to
the microwave field, so that our measurements are not affected by
detrimentals edge effects (unlike, e.g., dc and microwave
measurements on striplines).  Due to the field distribution of the
excited mode, the peak microwave field $\mu_{0} H_{\mu w}\sim$ 10
$\mathrm{\mu}$T is reached in an annular area of 4 mm diameter
centered on the center of the film.  Thus, for all practical purposes,
the amplitude of the microwave field $H_{\mu w}\ll H$, so all the
field effect should be regarded as coming from the dc applied field. 
Field-induced 
variations of the quality factor $\Delta\frac{1}{Q}$ and of the 
resonance frequency $\Delta\nu$ were measured at several temperatures 
in the range from 67 (69) to 90 K, resulting in reduced
temperature ranges above $T/T_{c}=$ 0.76 (0.79) in YBCO (SmBCO).  Data
have been collected either in zero-field-cooled, field-cooled, and on
direct and reverse field-sweeping.  We observed no hysteresis, with
the possible exception of the data below $\sim$70 K, where a small
hysteresis, smaller than 10\% of the overall signal is occasionally
present on reverse field sweeps.  In this case, data for field sweeps
after ZFC are considered.  We note that rather high operating
temperatures and the strong demagnetization present in our arrangement
(the dc field is perpendicular to the square film) lead to a very
small value of the first penetration field, and in particular to
$B\simeq\mu_{0}H$ even at low 
fields.\cite{demagnetization}  However,
for conservative reasons, we will not focus the discussion on the low 
field data (below 10 mT).
Figure \ref{fig1} reports the temperature 
variations of the cavity parameters reflecting the superconducting 
transition in zero magnetic field in YBCO (left panel) and SmBCO 
(right panel).  As can be seen, a substantial background is present in 
the measurement of the frequency shift, indicating relevant mechanical 
relaxations of the metallic resonator.  Even if the background was rather 
reproducible, we have chosen to work with field variations at fixed 
temperatures.  In this case, the variation of the effective impedance 
does not depend on the background (but only on a geometrical factor 
$G$, which acts as a scale factor), and $\Delta\frac{1}{Q}$ and 
$\Delta\nu$ directly yield the field induced changes of the effective 
surface impedance of the thin film, according to: 
$G\left[\Delta\frac{1}{Q}-\mathrm{i}2\frac{\Delta\nu}{\nu_{0}}\right]=\Delta 
Z_{s}^{eff}(H)\simeq\left[\Delta\tilde\rho(H)\right]/d=\left[\Delta\rho_{1}(H)+\mathrm{i} 
\Delta\rho_{2}(H)\right]/d$, where $\tilde\rho$ is the complex 
resistivity. Here, we have made use of the thin film approximation, 
extensively examined in Ref.\onlinecite{silvaSUSTthin}.\\
In figure \ref{fig2} we show the effect of the application of a 0.7 T 
magnetic field in YBCO and SmBCO, respectively.  The field-induced 
dissipation has a similar behaviour in both materials: 
$\left[\rho_{1}(T,0.7 \mathrm{T})-\rho_{1}(T,0)\right]/d$ increases 
with increasing temperature, and then vanishes at the transition 
temperature, as qualitatively expected. The variation of the imaginary 
part $\left[\rho_{2}(T,0.7 \mathrm{T})-\rho_{2}(T,0)\right]/d$ 
exhibits a quantitatively different behaviour in the two materials: in 
YBCO it is positive but small, apart the region very close to the 
transition where, at a temperature $T_{0}$, it becomes negative and 
eventually approaches zero for $T\rightarrow T_c$.  In SmBCO the 
qualitative behaviour is similar, with a larger low temperature value 
and a much lower $T_{0}$.  A deeper investigation is made by measuring 
the field-sweeps at fixed temperature.  In figures \ref{fig3} and 
\ref{fig4} we show typical field sweeps, 
$\Delta\tilde\rho(H)/d=\left[\tilde\rho(H,T)-\tilde\rho(0,T)\right]/d$, 
in YBCO and SmBCO, respectively, at temperatures chosen in both cases 
a few Kelvins below $T_{0}$.
The data here reported are representative of the behavior observed in
other SmBCO and YBaCuO films measured under the same conditions and in
similar temperature and field ranges. 
It is seen that the field dependences of the microwave resistivity in
YBCO and SmBCO are slightly different.  In particular,
in the present field range the field variation of both
$\Delta\rho_{1}$ and $\Delta\rho_{2}$ in SmBCO presents a more
pronounced downward curvature than in YBCO. We discuss in the
following the possibile reasons for the similarities and differences
between YBCO and SmBCO.

\section{Discussion}
\label{disc}

\begin{figure}
\centerline{
\psfig{figure=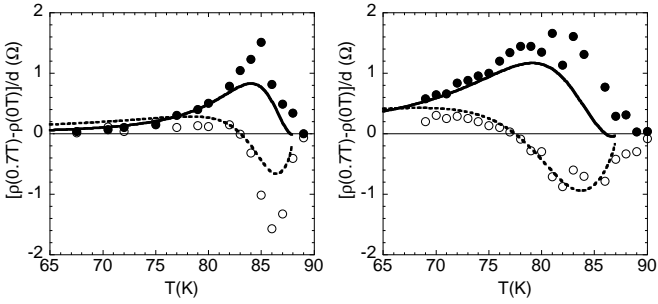,height=4cm,clip=,angle=0.}
}
\vspace{1cm}
\caption{Variation of the resistivity with the application of a 0.7 T 
magnetic field.  Left panel reports data on YBCO, right panel in 
SmBCO. Full symbols: shift of the real part, $\left[\rho_{1}(T,0.7 
{\mathrm T})-\rho_{1}(T,0)\right]/d$.  Open circles: shift in the 
imaginary part, $\left[\rho_{2}(T,0.7 {\mathrm 
T})-\rho_{2}(T,0)\right]/d$. Lines are calculated curves with 
Eq.\ref{rho2f} with a common scale factor as the only free parameter.}
\label{fig2}
\end{figure}

\begin{figure}
\centerline{
\psfig{figure=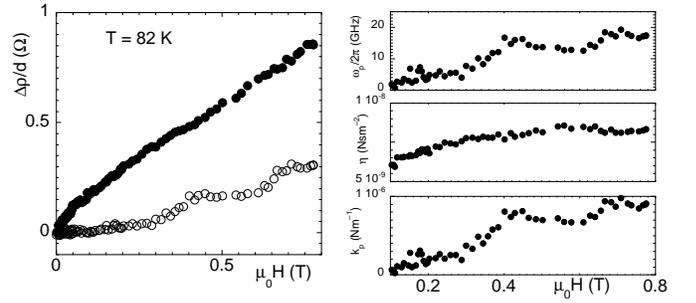,height=4cm,clip=,angle=0.}
}
\vspace{1cm}
\caption{Left panel: field induced change in the real (full circles) 
and imaginary (open circles) part of the complex resistivity.  Right 
panels: vortex parameters as calculated from data in YBCO accordingly 
to the conventional Gittleman-Rosenblum model.  From top to bottom: 
depinning frequency $\omega_{p}/2\pi$, unit length vortex viscosity $\eta$, 
unit length pinning constant $k_{p}$.  The strong field dependence of the 
so-calculated vortex viscosity and pinning constant cannot be easily 
justified.}
\label{fig3}
\end{figure}

\begin{figure}
\centerline{
\psfig{figure=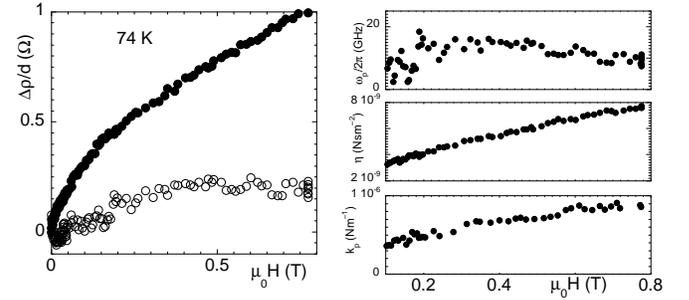,height=4cm,clip=,angle=0.}
}
\vspace{1cm}
\caption{Same as in Figure 3, but for the SmBCO film.}
\label{fig4}
\end{figure}

When we come to discuss the data, we are faced with the increase of the 
imaginary part in even moderate fields at low enough temperatures (see 
figures \ref{fig2},\ref{fig3},\ref{fig4}).  Since the role of pinning reveals 
itself mostly on the imaginary part, one might be tempted to assign 
the quantitative difference between YBCO and SmBCO to a different 
pinning, stronger in SmBCO. Following this quite conventional view, we 
first discuss the data in terms of simple vortex dynamics, assuming 
that the observed resistivity is entirely due to vortex motion (VM).  
To this aim, we take the simplest model for $\tilde\rho_{vm}$ as early 
proposed by Gittleman and Rosenblum\cite{gittle} (GR):
\begin{equation}
\label{GR}
\tilde{\rho} = \tilde\rho_{vm}(H,T)=
\frac{\Phi_{0}B}{\eta}
\frac{1+\mathrm{i}\frac{\omega_{p}}{\omega}}{1+\left(\frac{\omega_{p}}{\omega}\right)^{2}}
\end{equation}
where $\eta$ is the unit length vortex viscosity, $\omega_{p}$ is the
depinning angular frequency and $k_{p}=\omega_{p}\eta$ is the unit
length pinning constant.  This analysis has been widely used in
various HTCS, see the review in Ref.\onlinecite{golos}.  This model,
based on the motion of independent vortices, is expected to be more
and more accurate with increasing frequency: at high operating
frequencies vortex displacements induced by the microwave field are
much smaller than the intervortex mean spacing, and complications
arising from the collective nature of the vortex matter are less and
less relevant.  In fact, it has been experimentally shown\cite{wu}
that in YBCO a dynamic crossover from a glassy behaviour to the
independent vortex motion takes place above a few GHz.  We note that
the GR model explicitly predicts that the imaginary resistivity should
{\it increase} with the application of the field.  This result is
common to other VM based models, based on less restrictive
assumptions.  It is then apparent that, at least in a temperature
range of several Kelvins below $T_{c}$ (above $T_{0}$), the data
cannot be explained on the basis of VM alone.  At low enough
temperatures, applying in a straightforward way the GR model one would
get directly from the data the depinning frequency $\omega_{p}/2\pi$,
the vortex viscosity $\eta$ and the pinning constant $k_{p}$.  In
figures \ref{fig3} and \ref{fig4} (right panels) we report the
calculated vortex parameters in YBCO and SmBCO. We immediately note
that in both materials the vortex viscosity and pinning constant
increase with the field, while the depinning frequency increases in
YBCO and is approximately constant in SmBCO. The field dependence of
the so-obtained $\eta$ and $k_{p}$ is not easily explained.  In
particular, one would expect a nearly constant vortex viscosity, as
experimentally determined by multifrequency measurements at lower
temperatures\cite{tsuchiya} and by high-field swept-frequency Corbino
disk measurements,\cite{sartiPhC} and a weak field dependence of
$k_{p}$.  Exotic field dependences of the vortex viscosity, increasing
with the field as $\eta\sim\sqrt{B}$, have been observed in
Bi$_{2}$Sr$_{2}$CuO$_{6}$ and tentatively explained \cite{matsuda} in
terms of a different field dependence of quasiparticle relaxation time
in a d-wave superconductor, appearing at high fields when the
intervortex distance becomes smaller than the mean free path.  This
picture does not apply easily to our case, since the field range is
very different.  In addition, this interpretation would not give an
explanation of the field increase of $k_{p}$ (we remark that the
anomalies reported in Ref.\onlinecite{matsuda} were confined to $\eta$:
$k_{p}$ was found constant at low fields, and decreasing at higher
fields).  We conclude that this explanation is at least questionable
at low fields.\\
Granularity is sometimes indicated as a possible dominant source for
the losses in the microwave response in superconducting films.
Manifestations of granularity include weak-links
dephasing,\cite{giura} Josephson fluxon (JF) dynamics
\cite{halbritter} and, as recently studied, Abrikosov-Josephson fluxon
(AJF) dynamics.\cite{gurevich}  Weak-links dephasing is charaterized
by a very sharp increase of the dissipation at dc fields of order or
less than 20 mT, accompanied by a strong (and sometimes exceptionally
strong) hysteresis with increasing or decreasing field.\cite{giura}
However, those effects are relevant in large-angle grain boundaries,
such as those found in pellets and granular samples, and are not
observed in good thin films.  In fact, we did not observe none of the
above mentioned effects in our films, so we conclude that weak-links
dephasing does not affect our measurements.
Josephson fluxon dynamics
has been studied essentially in relation to nonlinear effects, due to
the short JF nucleation time. 
If however one assumes that a dc field has the same effect as a
microwave field, the qualitative properties of JF dynamics would be
barely distinguishable from conventional Abrikosov fluxons dynamics,
since an equation like Eq.\ref{GR} would hold,\cite{halbritter} with
the noticeable difference of a larger pinning frequency (due to the
small viscosity, due in its turn to the insulating core).  Reactive
response should then dominate, while in our measurements (see figs.
\ref{fig2},\ref{fig3},\ref{fig4}) the dissipation ($\Delta\rho_{1}$)
is always larger than reactance ($\Delta\rho_{2}$).
We are led to 
conclude that JF dynamics, even if possibly present, is not the main 
mechanism driving the measured microwave response.\\
Another possibility is the role played by Abrikosov-Josephson fluxons.
Such flux structures nucleate along small-angle grain boundaries, as
those presumably present in our oriented films.  The ac response is
expected to saturate at fields larger than a characteristic field
$H_{0}$, whose estimate spans orders of magnitude,\cite{gurevich} in
the range 0.1-10 T. In YBCO bicrystals, dc measurements\cite{gurexp}
yielded $H_{0} < $ 2 T above $T=$ 70 K. However, it is noteworthy that
in  defective thin films of other cuprates (Tl$_{2}$Ba$_{2}$CaCu$_{2}$O$_{8+x}$)
the 8.5 GHz dissipation ascribed to AJ fluxon motion is shown to
saturate at (microwave) fields as low as $\mu_{0}H_{0}\sim$ 1
mT,\cite{gaganidze} while in
better films this kind of response is not present at all.
Nevertheless, allowing for a large
$H_{0}$, the explicit expression for AJ fluxon motion complex
resistivity\cite{gurevich} yields an initial magnetic-field increase
in the resistivity as $\sim\sqrt{\frac{H}{H_{0}}}$, whereas the
reactance initially decreases with the magnetic field as
$\sim-\sqrt{\frac{H}{H_{0}}}+c(T)\frac{H}{H_{0}}$ with
$-\frac{1}{2}<c<\frac{1}{2}$.  It appears that our data, where the
reactance increases or decreases with the field dependending on the
temperature, are not easily described by this only mechanism.\\
Summarizing, none of the mechanisms shortly recalled above appears to
describe the main qualitative features of our data, namely: {\it (i)}
the sublinear variation of the real and imaginary parts of the
resistivity, {\it (ii)} the change of sign of the variation of the
reactance with increasing temperature.  It is not excluded that a
properly tailored combination of several of the above mechanisms could
describe our data.  Nevertheless, in order to look for a single
possibly dominant mechanism in the microwave complex response to a dc
magnetic field
we now comment on a possible alternative explanation for our
data. In fact there is quite robust evidence, in other HTCS, that in
the vortex state the quasiparticle increase has a relevant role, at 
least in the
imaginary response.\cite{mallozzi}  This phenomenon is well explained
in terms of a semiclassical theory of a superconductor with line of
nodes in the gap \cite{volovik}.  In particular, the decrease of
superfluid fraction is predicted to follow a $B^{\alpha}$ law, with
$\alpha=\frac{1}{2}$ in clean d-wave superconductors. In this case the
charge carrier conductivity can be described by the two-fluid model,
and cast in the form:
\begin{equation}
\label{sigma}
\tilde\sigma(H,T)= \sigma_{R0} x_{n}(T,B)-{\mathrm i} \sigma_{I0} x_{s}(T,B)
\end{equation}
where $x_{n}(T,B)$ and $x_{s}(T,B)=1-x_{n}(T,B)$ are the
temperature and field dependent normal and superfluid fractional
densities, and
$x_{s}(T,B)=x_{s}(T,0)\left[1-(B/B_{pb})^{\alpha}\right]$ where $B_{pb}$
is a pair breaking field.  It has been shown\cite{won} that in clean
superconductors $B_{pb}\propto B_{c2}(T)$, but deviations from this
behaviour are possible.\\
As a very crude approximation, we can assume as a limiting case the 
vortex motion does not give a significant contribution to the 
resistivity. Taking into account that $(B/B_{pb})^{\alpha}\ll 1$, one 
can show\cite{silvaSmBCO} that in this case the field variation of the complex 
resistivity can be written as:
\begin{equation}
\label{rho2f}
\Delta\tilde\rho(H,T)=\left[a_{1}(T;s)+{\mathrm i}
a_{2}(T;s)\right]\left(\frac{B}{B_{pb}}\right)^{\alpha}
\end{equation}
where the dependence on the single parameter
$s=\frac{\sigma_{R0}}{\sigma_{I0}}$ has been made explicit.  By
considering the temperature dependent data in figure \ref{fig2}, we do
not need to assume any specific field dependence.  However, we remark
that in SmBCO the data are well described with $\alpha=\frac{1}{2}$
(Ref.\cite{silvaSmBCO}).  Due to the oversimplification of the model,
for the temperature dependences involved in Equation \ref{rho2f} we
take conservatively the most simple: the superfluid fraction
$x_{s}(T,0)=\left(1-t^{2}\right)$, and $B_{pb}\propto B_{c2}(T)$, so
that $B_{pb}=B_{pb0}\left(1-t^{2}\right)$, with $t=T/T_{c}$.  In order
to gain qualitative information, we do not attempt to insert some
temperature dependence in the parameter $s$.  We stress that, with
this choice, $s$ is not a free parameter: its value is determined by
the requirement that the calculated curve of $\Delta\rho_{2}$ changes
sign at the experimental temperature $T_{o}$.  The only free parameter
is the scale factor given by the pair breaking field.  The model is
surely oversimplified, but has rather strong constraints: in
particular the shapes of the two calculated curves $\Delta\rho_{1}(T)$
and $\Delta\rho_{2}(T)$ are determined, and only a common scale factor
can be adjusted.  The resulting calculated curves for the
experimentally applied field $B_{a}=$ 0.7 T are reported in figure
\ref{fig2}.  As can be seen, the overall shape of the experimental
curves are reproduced.  From the scale factors we estimate
$\left(\frac{B_{a}}{B_{pb0}}\right)^{\alpha}\sim$ 0.05 in YBCO and
SmBCO. One also gets $s\simeq$ 0.1 and 0.2 in YBCO and SmBCO,
respectively.  We note that by applying the conventional two-fluid
model, $\tilde\sigma= \frac{ne^{2}}{m\omega}\left[\omega\tau
x_{n}-{\mathrm i} x_{s}\right]$, one has $s=\omega\tau$ where $\tau$
is the quasiparticle scattering time.  Taking into account the
measuring frequency, one has $\tau\simeq$ 0.35 ps and 0.7 ps in YBCO
and SmBCO, respectively, that compare well to, e.g., 0.2 ps at
$\approx$80 K as obtained from microwave measurements in YBCO
crystals\cite{bonnPRB93} and to 0.5 ps at $\approx$80 K as obtained
from millimeter-wave interferometry in YBCO film.\cite{nagashima} We
note that the model appears to describe better the data on SmBCO than
on YBCO: this is in fact consistent with the much stronger curvature
of the field-sweeps data in SmBCO, since the proposed field-induced
pair breaking mechanism is naturally sublinear in superconductors with
nodes in the gap.  In fact, in SmBCO the small difference between the
calculated pair breaking response and the experimental data can be
well described by vortex motion in the free-flow
limit.\cite{silvaSmBCO,silvaPhC} In this respect, the discrepancy
between the calculated pair-breaking and the experimental data, as
well as deviations from $H^{\frac{1}{2}}$ dependence of the field
sweeps, is most likely due to motion of flux lines, either JF, AJF or
Abrikosov fluxons.
We believe that the successful application of the oversimplified 
pair-breaking
model in the qualitative description of the vortex-state resistivity
in RE(Y,Sm)BCO points toward a substantial contribution of pair
breaking by the magnetic field, in agreement with the existence of
lines of nodes in the gap,
while additional mechanisms, such as fluxon motion, are most probably 
also present.

\section{Conclusion}
\label{conc}

We have presented data for the microwave resistivity in 
REBa$_{2}$Cu$_{3}$O$_{7-\delta}$ thin films as a function of the 
temperature and magnetic field.  The field-dependence of the complex 
resistivity exhibits quantitatively different behaviours in the two 
materials.  The analysis in terms of vortex motion alone, in 
particular ascribing the quantitative differences to a different 
pinning, would imply very exotic field dependences 
of the vortex parameters, so that it appears that alternative 
interpretations have to be found, including a possible enhanced role 
of the field-induced pair breaking.  By ascribing, as a limiting case, 
the entire response to the field-induced pair-braking we obtain 
semi-quantitative fits of the data at 0.7 T as a function of the 
temperature, with a large quasiparticle scattering time below $T_{c}$ 
in agreement with estimates given by several groups.  Additional 
measurements might be required in order to fully understand the 
interplay between fluxon motion and pair breaking.

\section*{acknowledgements}

Valuable assistance by L. Muzzi is warmly
acknowledged.  We benefited from useful discussions and remarks by
M.W. Coffey, J. R. Clem, S. Anlage, D. Oates, R. Woerdenweber, 
J. Halbritter and A.
Maeda.  We thank A.M. Cucolo and M. Boffa at Universit\`a di Salerno
for the SmBCO sample.  This work has been partially supported by
Italian MIUR under FIRB ``Strutture semiconduttore/superconduttore per
l'elettronica integrata''.

\newpage

\end{multicols}

\end{document}